\documentclass[10pt,journal,cspaper,compsoc]{IEEEtran}
\ifCLASSOPTIONcompsoc
   \usepackage[nocompress]{cite}
\else
\fi
\ifCLASSINFOpdf
\else
   \usepackage[dvips]{graphicx}
\fi
\usepackage[cmex10]{amsmath}
\usepackage{amssymb}
\usepackage{amsthm}

\begin{document}
%
% paper title
% can use linebreaks \\ within to get better formatting as desired
 \title{Space-efficient Verifiable Secret Sharing Using Polynomial Interpolation}
%
%
% author names and IEEE memberships
% note positions of commas and nonbreaking spaces ( ~ ) LaTeX will not break
% a structure at a ~ so this keeps an author's name from being broken across
% two lines.
% use \thanks{} to gain access to the first footnote area
% a separate \thanks must be used for each paragraph as LaTeX2e's \thanks
% was not built to handle multiple paragraphs
%
%
%\IEEEcompsocitemizethanks is a special \thanks that produces the bulleted
% lists the Computer Society journals use for "first footnote" author
% affiliations. Use \IEEEcompsocthanksitem which works much like \item
% for each affiliation group. When not in compsoc mode,
% \IEEEcompsocitemizethanks becomes like \thanks and
% \IEEEcompsocthanksitem becomes a line break with idention. This
% facilitates dual compilation, although admittedly the differences in the
% desired content of \author between the different types of papers makes a
% one-size-fits-all approach a daunting prospect. For instance, compsoc 
% journal papers have the author affiliations above the "Manuscript
% received ..."  text while in non-compsoc journals this is reversed. Sigh.

\author{\IEEEauthorblockN{Massimo~Cafaro*,~\IEEEmembership{Senior~Member,~IEEE}
        and~Piergiuseppe~Pell\`e}% <-this % stops an unwanted space
\IEEEcompsocitemizethanks{\IEEEcompsocthanksitem The authors are with the Department of Engineering for Innovation, University of Salento, Lecce 73100, Italy \protect\\     
% note need leading \protect in front of \\ to get a newline within \thanks as
% \\ is fragile and will error, could use \hfil\break instead.
E-mail: massimo.cafaro@unisalento.it, piergiuseppe@gmail.com}
\thanks{}}

% note the % following the last \IEEEmembership and also \thanks - 
% these prevent an unwanted space from occurring between the last author name
% and the end of the author line. i.e., if you had this:
% 
% \author{....lastname \thanks{...} \thanks{...} }
%                     ^------------^------------^----Do not want these spaces!
%
% a space would be appended to the last name and could cause every name on that
% line to be shifted left slightly. This is one of those "LaTeX things". For
% instance, "\textbf{A} \textbf{B}" will typeset as "A B" not "AB". To get
% "AB" then you have to do: "\textbf{A}\textbf{B}"
% \thanks is no different in this regard, so shield the last } of each \thanks
% that ends a line with a % and do not let a space in before the next \thanks.
% Spaces after \IEEEmembership other than the last one are OK (and needed) as
% you are supposed to have spaces between the names. For what it is worth,
% this is a minor point as most people would not even notice if the said evil
% space somehow managed to creep in.

% The paper headers
\markboth{Journal of \LaTeX\ Class Files,~Vol.~6, No.~1, January~2007}%
{Cafaro \MakeLowercase{\emph{et al.}}: Space-efficient Verifiable Secret Sharing Using Polynomial Interpolation}
% The only time the second header will appear is for the odd numbered pages
% after the title page when using the twoside option.
% 
% *** Note that you probably will NOT want to include the author's ***
% *** name in the headers of peer review papers.                   ***
% You can use \ifCLASSOPTIONpeerreview for conditional compilation here if
% you desire.

% The publisher's ID mark at the bottom of the page is less important with
% Computer Society journal papers as those publications place the marks
% outside of the main text columns and, therefore, unlike regular IEEE
% journals, the available text space is not reduced by their presence.
% If you want to put a publisher's ID mark on the page you can do it like
% this:
%\IEEEpubid{0000--0000/00\$00.00~\copyright~2007 IEEE}
% or like this to get the Computer Society new two part style.
%\IEEEpubid{\makebox[\columnwidth]{\hfill 0000--0000/00/\$00.00~\copyright~2007 IEEE}%
%\hspace{\columnsep}\makebox[\columnwidth]{Published by the IEEE Computer Society\hfill}}
% Remember, if you use this you must call \IEEEpubidadjcol in the second
% column for its text to clear the IEEEpubid mark (Computer Society jorunal
% papers don't need this extra clearance.)

% for Computer Society papers, we must declare the abstract and index terms
% PRIOR to the title within the \IEEEcompsoctitleabstractindextext IEEEtran
% command as these need to go into the title area created by \maketitle.
\IEEEcompsoctitleabstractindextext{%
\begin{abstract}
Preserving data confidentiality in clouds is a key issue. Secret Sharing, a cryptographic primitive for the distribution of a secret among a group of $n$ participants designed so that only subsets of shareholders of cardinality $0 < t \leq n$ are allowed to reconstruct the secret by pooling their shares, can help mitigating and minimizing the problem. A desirable feature of Secret Sharing schemes is cheater detection, i.e. the ability to detect one or more malicious shareholders trying to reconstruct the secret by obtaining legal shares from the other shareholders while providing them with fake shares. Verifiable Secret Sharing schemes solve this problem by allowing shareholders verifying the others' shares. We present new verification algorithms  providing arbitrary secret sharing schemes with cheater detection capabilities, and prove their space efficiency with regard to other schemes appeared in the literature. We also introduce, in one of our schemes, the Exponentiating Polynomial Root Problem (EPRP), which is believed to be NP-Intermediate and therefore difficult. 
\end{abstract}

% IEEEtran.cls defaults to using nonbold math in the Abstract.
% This preserves the distinction between vectors and scalars. However,
% if the journal you are submitting to favors bold math in the abstract,
% then you can use LaTeX's standard command \boldmath at the very start
% of the abstract to achieve this. Many IEEE journals frown on math
% in the abstract anyway. In particular, the Computer Society does
% not want either math or citations to appear in the abstract.

% Note that keywords are not normally used for peer review papers.
}

% make the title area
\maketitle

% To allow for easy dual compilation without having to reenter the
% abstract/keywords data, the \IEEEcompsoctitleabstractindextext text will
% not be used in maketitle, but will appear (i.e., to be "transported")
% here as \IEEEdisplaynotcompsoctitleabstractindextext when compsoc mode
% is not selected <OR> if conference mode is selected - because compsoc
% conference papers position the abstract like regular (non-compsoc)
% papers do!
\IEEEdisplaynotcompsoctitleabstractindextext
% \IEEEdisplaynotcompsoctitleabstractindextext has no effect when using
% compsoc under a non-conference mode.

% For peer review papers, you can put extra information on the cover
% page as needed:
% \ifCLASSOPTIONpeerreview
% \begin{center} \bfseries EDICS Category: 3-BBND \end{center}
% \fi
%
% For peerreview papers, this IEEEtran command inserts a page break and
% creates the second title. It will be ignored for other modes.
\IEEEpeerreviewmaketitle

\section{Introduction}
% Computer Society journal papers do something a tad strange with the very
% first section heading (almost always called "Introduction"). They place it
% ABOVE the main text! IEEEtran.cls currently does not do this for you.
% However, You can achieve this effect by making LaTeX jump through some
% hoops via something like:
%
%\ifCLASSOPTIONcompsoc
%  \noindent\raisebox{2\baselineskip}[0pt][0pt]%
%  {\parbox{\columnwidth}{\section{Introduction}\label{sec:introduction}%
%  \global\everypar=\everypar}}%
%  \vspace{-1\baselineskip}\vspace{-\parskip}\par
%\else
%  \section{Introduction}\label{sec:introduction}\par
%\fi
%
% Admittedly, this is a hack and may well be fragile, but seems to do the
% trick for me. Note the need to keep any \label that may be used right
% after \section in the above as the hack puts \section within a raised box.

% The very first letter is a 2 line initial drop letter followed
% by the rest of the first word in caps (small caps for compsoc).
% 
% form to use if the first word consists of a single letter:
% \IEEEPARstart{A}{demo} file is ....
% 
% form to use if you need the single drop letter followed by
% normal text (unknown if ever used by IEEE):
% \IEEEPARstart{A}{}demo file is ....
% 
% Some journals put the first two words in caps:
% \IEEEPARstart{T}{his demo} file is ....
% 
% Here we have the typical use of a "T" for an initial drop letter
% and "HIS" in caps to complete the first word.

\newtheorem{thm}{Theorem}[section]
\newtheorem{lem}[thm]{Lemma}
\newtheorem{prop}[thm]{Proposition}
\newtheorem*{cor}{Corollary}

\newtheorem{defn}{Definition}[section]
\newtheorem{conj}{Conjecture}[section]
\newtheorem{exmp}{Example}[section]

\newtheorem*{rem}{Remark}
\newtheorem*{note}{Note}
\newtheorem{case}{Case}

\IEEEPARstart{S}{ecret}  Sharing deals with the problem of securely distributing confidential information among a certain number of shareholders, in such a way that only some subsets of them are able to jointly decrypt it. Several schemes and variants of secret sharing have been proposed, from the seminal schemes of Shamir  \cite{journals/cacm/Shamir79} and Blakley  \cite{Blakley1979}, which are based respectively on polynomial interpolation, and hyperplanes intersection, to the newest approaches closely involving number theory, such as the ones based on the Chinese Remainder Theorem  \cite{Mignotte} \cite{Asmuth}. \\Secret Sharing can be beneficial in many different ways in cloud computing, which is becoming increasingly common, with rapid adoption by both industry, small and medium enterprises, and individual users. Among the many services provided by a cloud infrastructure, we are concerned here with cloud storage and file hosting services. Building on a highly virtualized infrastructure, these services are succeeding owing to economic reasons and to the fact that the underlying infrastructure and physical location are fully transparent to the user. However, preserving data confidentiality in clouds is a key issue  \cite{Tysowski}.
The main difficulty is related to the fact that data is stored on a remote server which is fully accessible by the cloud service provider (and can be accessible to third-party people through a malicious attack). In order to achieve data confidentiality and to overcome this issue, it is possible to encrypt a file containing sensitive information before storing it on a cloud. Even though encryption makes harder unauthorized disclosure of information, a better solution is based on the use of Secret Sharing and multiple cloud providers, a scenario in which each generated share is stored on a different cloud. The original file can still be encrypted if required, thus providing an additional security guarantee. The use of multiple clouds and Secret Sharing can therefore mitigate and minimize several risks associated to the single cloud provider scenario, such as service availability failure, data loss and/or corruption, loss of confidentiality, vendor lock-in and the possibility of malicious insiders in the single cloud.\\
One important issue in the design of a secret sharing protocol is its robustness against cheaters: common solutions proposed in the literature rely on checking consistency of the secret information after reconstruction from more than one group of shareholders, or on adding helpful data to the shares in order to detect and/or identify mistrustful behaviour. Verifiable Secret Sharing (VSS) \cite{Chor} is therefore secret sharing augmented with features that allow only detection or also identification of any cheater in a coalition, unconditionally or with respect to the scheme parameters (threshold value, total number of dishonest
shareholders, etc.). Several VSS schemes have been proposed, including, for instance, Publicly Verifiable Secret Sharing (PVSS) \cite{Stadler}  \cite{Fujisaki-Okamoto}  \cite{Schoenmakers}  \cite{Boudot} \cite{Peng-Bao}  \cite{Peng} or schemes focusing on Asynchronous Verifiable Secret Sharing (AVSS) such as \cite{Beerliov} \cite{Ben-Or} \cite{Canetti95} \cite{Ben-Or:1994} \cite{Canetti:1993} \cite{Patra} \cite{Patra2015}.\\ In this work, we present new verification algorithms based on commitments providing arbitrary secret sharing schemes with cheater detection capabilities, and prove their data efficiency with regard to other schemes appeared in the literature. Our approach belongs to the Honest-Dealer VSS scheme category \cite{Cramer2001}\cite{Rabin1994}, since it requires a one-time honest dealer. Our contribution is three-fold: (i) we present space-efficient verification protocols that does not even require storing public data for verification; (ii) our schemes can be used in conjunction with arbitrary secret sharing schemes, and provide cheater detection capabilities; (iii) we also introduce, in one of our schemes, a new computational problem, namely the Exponentiating Polynomial Root Problem (EPRP), which generalizes the Discrete Logarithm Problem (DLP).\\ 
The remainder of this paper is organized as follows. Section \ref{related-work-section} recalls related work. We present our space-efficient verifiable schemes and analyze their security in  Section \ref{schemes-section}, along with EPRP. In Section \ref{runtime-efficiency-section}, we propose runtime efficiency refinements to optimize our schemes. The information rates of our schemes are discussed in Section \ref{comparison-section}, in which we also compare our schemes against the state of the art schemes published in the literature. Finally, we draw our conclusions and propose future work in Section \ref{conclusions-section}.

\section{Related Work}
\label{related-work-section}

In this Section, we discuss related work. We begin by reviewing commitments, and then proceed analyzing hashing, the schemes based on homomorphic commitments proposed by Feldman \cite{Feldman}, Pedersen \cite{conf/crypto/Pedersen91} and Benaloh \cite{conf/crypto/Benaloh86a} and the set coherence verification method, introduced by Harn and Lin \cite{journals/dcc/HarnL09}.

A commitment  \cite{Blum1983} is a statement that proves knowledge of some information, \emph{without revealing the information itself}. 
A formal definition follows:

\begin{defn}[Commitment]
Given a value $x$, a commitment $c(x)$ is a value such that the following conditions are satisfied:
\begin{itemize}
\item \textbf{Hiding:} By knowledge of $c(x)$, it is impossible (or very difficult) to obtain $x$ --- $c(x)$ \emph{hides} $x$;
\item \textbf{Binding:} It is infeasible or impossible to find another value $y$ for which $c(y)=c(x)$ --- $c(x)$ \emph{binds} to $x$.
\end{itemize}
\end{defn}

The two properties just defined may refer to the \emph{computational} or to the \emph{unconditional security} setting: if an attacker with infinite computing power can break the former or the latter, the scheme is said to be, respectively, \emph{computationally hiding} or \emph{computationally binding}. Otherwise, a commitment scheme is said to be \emph{unconditionally hiding} or \emph{unconditionally binding}. More precisely, it can be proved \cite{books/sp/DelfsK02} that a commitment scheme cannot be simultaneously \emph{unconditionally hiding} and \emph{unconditionally binding}. Commitments can be implemented via one-way functions, as a basis for verification schemes.

\subsection{Hashing}
The simplest method to add verification capabilities to a scheme, is to use one-way functions to obtain fingerprints/signatures
of the data involved.
Two trivial algorithms for detection and identification are listed (suppose that $H$ is a secure hash function):\\

\textbf{Detection}\\
Dealer: Given the secret $s$, compute $h = H(s)$ and make it public.\\
Shareholder: After reconstructing a secret $x$, verify whether $H(x) = h$. If $H(x) \neq h$ someone is cheating.\\

\textbf{Identification}\\
Dealer: Given the shares $s_1,\ldots,s_n$, compute the signatures $h_i = H(s_i)$ for every $i$ and make them public.\\
Shareholder: Before performing reconstruction, for every share $s_j$ received, get $h_j$ and check that ${H(s_j) = h_j}$. If equality
does not hold, then shareholder $j$ is cheating.

The clear disadvantage of identification by hashing is that verification data \emph{grows linearly} with $n$.

\subsection{Homomorphic commitments: Feldman's scheme}
\label{feldmanvss}
Feldman's scheme \cite{Feldman} is a verification method applicable to Shamir's secret sharing.
Like hashing, it relies on the use of one-way functions for verifying
consistency of each share. Moreover, the homomorphic property is exploited in order to decrease
the total number of verification elements from $n$ to $t$ -- the commitment is over the secret, not over the shares.
Indeed, let $v$ be a $(+,\cdot)$-homomorphic
one-way function (that is, $v(a+b)=v(a)v(b)$); then, if $v$ is evaluated over a polynomial, the following
equation holds:
\begin{equation}
v\left( {\mathop \sum \limits_{i = 0}^{t - 1} {a_i}{x^i}} \right) = \mathop \prod \limits_{i = 0}^{t - 1} v\left( {a_i^{{x^i}}} \right)
\end{equation}
The scheme steps are reported below:

\begin{itemize}
\item{Choose as public values primes $p,q$ such that $q$ divides $p-1$ and a generator $\alpha$ of a subgroup of order $q$ of $\mathbb{Z}_p^*$ ($q$ is the lowest possible integer such that $\alpha ^ q \equiv 1 \mod p$)}; the bitsize of $q$ is much lower than the one of $p$, and this is done since not only finding primitive roots, but even computing multiplicative orders for generic moduli, are, in general, hard problems (random sampling and factorization of the modulus are used for better efficiency). Theoretically, one could also choose a generator of order $p-1$;
\item{Starting with the secret $a_0$, generate the polynomial:
$$P(x)=a_0 + \ldots + a_{t-1}x^{t-1}$$
over the field $\mathbb{Z}_q$, from which the shares are sampled as $s_i = P(i)$, $i=1,\ldots,n$};
\item{Generate the public verification coefficients:
$$\alpha_j = \alpha^{a_j} \mod p,~{j=0,\ldots,t-1}$$}
\item{Thanks to the homomorphic property of exponentiation, a commitment to a share $s_i$ can be written as:

\begin{IEEEeqnarray}{rCl}
\alpha^{s_i} & = & \alpha^{P(i)} = \alpha^{a_0 + a_1 i + \ldots + a_{t-1} i^{t-1}} \IEEEnonumber\\
&&=\ \alpha^{a_0} {\alpha^{a_1}}^i \ldots {\alpha^{a_{t-1}}}^{i^{t-1}}  \IEEEnonumber\\
&&=\ \alpha_0 \alpha_1^i \ldots \alpha_{t-1}^{i^{t-1}}%
\end{IEEEeqnarray}

Hence, the consistency of a share $s_i$ can be verified by checking the equality: 

\begin{equation}
\alpha^{s_i} \equiv \mathop \prod \limits_{j = 0}^{t - 1} \alpha _j^{i^j} \pmod p
\end{equation}
}

\end{itemize}

It is worth noting here that the one-way function candidate used here is modular exponentiation over
$\mathbb{Z}_p^*$.

Feldman's scheme is computationally hiding, since exponentiation is done over the secret polynomial's coefficients, so solving the DLP would allow to obtain the secret from the verification data (reverse hiding). It is also unconditionally binding since the mapping between values and commitments is injective, so multiple values committing to the same output cannot be found.

\subsection{Homomorphic commitments: Pedersen's scheme}
\label{pedersenvss}
With some slight modifications, proposed in \cite{conf/crypto/Pedersen91}, the previous scheme can be made perfectly hiding and computationally binding -- notice also that information rate grows, as there is more data to provide shareholders with.

\begin{itemize}
\item Choose as public parameters primes $p$ and $q$ as before, together with \emph{two} generators of order $q$, namely $g,h$;
\item Let $y(x) = a_0 + a_1 x + \ldots + a_{t-1} x^{t-1}$ be the polynomial to be committed. Generate an additional polynomial $z(x)$ of the same degree, with random non-null coefficients $b_0,\ldots,b_{t-1}$;
\item Compute the coefficients commitments as ${c_i = g^{a_i} h^{b_i} \mod p}$ and send them to every shareholder;
\item Sample the points for shareholder $j$ as ${y_j=y(j),~z_j=z(j)}$, then the share for shareholder $j$ is $\left(y_j,z_j\right)$;
\item As in the previous scheme, by applying the homomorphic property, a commitment to a share $\left(y_j,z_j\right)$ can be expressed as:

\begin{IEEEeqnarray}{rCl}
 g^{y_j}h^{z_j} &=&  g^{\mathop \sum \limits_{i = 0}^{t - 1} a_i j^i} h^{\mathop \sum \limits_{i = 0}^{t - 1} b_i j^i} \IEEEnonumber\\
&&=\ {\mathop \prod \limits_{i = 0}^{t - 1} {\left(g^{a_i}\right)}^{j^i}} \cdot {\mathop \prod \limits_{i = 0}^{t - 1} {\left(h^{b_i}\right)}^{j^i}} \IEEEnonumber\\
&&=\ {\mathop \prod \limits_{i = 0}^{t - 1} {\left( g^{a_i} h^{b_i}\right)}^{j^i}}  = {\mathop \prod \limits_{i = 0}^{t - 1} {c_i}^{j^i}}%
\end{IEEEeqnarray}

Thus, any shareholder can verify that a share $(y_j,z_j)$ is valid, by checking the equation:

\begin{equation}
g^{y_j}h^{z_j} \equiv \mathop \prod \limits_{i = 0}^{t - 1} {c_i}^{j^i} \pmod p
\end{equation}

\end{itemize}

Perfect hiding for a commitment $g^a h^b$ means that, for any triple $a,b,a'$, a value $b'$ exists such that ${g^a h^b \equiv g^{a'} h^{b'} \pmod p}$. This can be seen by expressing $h$ as a power of $g$: ${h = g^w \mod p}$, and it can always be done since ${h \in \mathbb{Z}_p^*}$. A commitment can then be expressed as:

\begin{equation}
c = g^{a+wb} \mod p
\end{equation}

Hence, by fixing the triple defined before, $b'$ can be found by solving:

\begin{equation}
a+wb \equiv a' + wb' \pmod p
\end{equation}

which is always well-defined.

\subsection{Homomorphic commitments: Benaloh's scheme}
\label{benalohvss}
This scheme \cite{conf/crypto/Benaloh86a} allows shareholders verifying that all of the shares are collectively $t$--consistent (i.e., an arbitrary subset $t$ of $n$ shares yields the same, correct, polynomial without revealing the secret). Verification is done through homomorphic algebra, without exposing the secret. However, the scheme requires an interactive proof to prove the dealer's integrity, which has been avoided by design in our scheme. Moreover, the proof involves the generation and use of a very large number of polynomials of degree $t$ for a $(t, n)$ threshold scheme, making the scheme impractical.

\subsection{Verifiability by set coherence}
\label{setcoherence}
This method, introduced in \cite{journals/dcc/HarnL09}, does not require any additional verification data besides the shares themselves.
However, when applied to a $(t,n)$-threshold scheme, it needs a coalition consisting
of $m$ shareholders, $m>t$. Cheater detection and identification are performed by comparing the secrets reconstructed by all of the
possible subsets of $t$ out of $m$ shareholders. The two algorithms follow.

\textbf{Detection}

\begin{itemize}
\item{Let $B$ be an authorized subset of size $m>t$ for a $(t,n)$-threshold scheme. For every subset
$A \subset B$ of size $t$, run the reconstruction algorithm with the corresponding shares. Keep a
histogram of all of the secrets found};
\item{If every subset $A\subset B$ rebuilds the same secret, there is no cheating. Otherwise, run the cheater identification algorithm.}
\end{itemize}

\textbf{Identification}

\begin{itemize}
\item{Select the \emph{majority secret} $s_m$ as the one with the highest frequency in the histogram. Assume it to be the
\emph{actual secret} (remember that this requires a honest majority). Take a subset $A$
that rebuilds $s_m$ (this can be done in constant time, if the histogram structure
keeps track of which subsets rebuild each secret)};
\item{Let $A = \{1,2,\ldots,t\}$ without loss of generality. Since $A$ rebuilds the correct secret
by assumption, then every share in $A$ is posted by a honest shareholder, and every possible cheater must be contained in
$C = B/A$};
\item{For every shareholder $j \in C$, check whether the set ${A'=\{j,2,\ldots,t\}}$
rebuilds $s_m$. If it does not, add shareholder $j$ to the cheater's list; again, this can be done in constant time, using the
augmented histogram of first step.}
\end{itemize}

Distinguishing between independent cheaters and organized ones, the bounds for detection and identification are summarized
in Table \ref{table2} ($c$ denotes the number of cheaters, $m$ the cardinality of $B$, $t$ the threshold value).
    
    \begin{table}
    \renewcommand{\arraystretch}{1.3}
     \caption{Set coherence: bounds for detection and identification}
\label{table2}
\centering
    \begin{tabular}{l|l|l}
    \hline
    ~              & \bfseries Independent cheaters &  \bfseries Organized cheaters \\ 
    \hline
    \hline
  \bfseries  Detection      & $m>t$                & $m-c>t$            \\ \hline
 \bfseries   Identification & $m-c>t$              & $m-c \geq c+t$     \\ \hline
    \end{tabular}
    \end{table}

\begin{rem}
Besides requesting a higher threshold value for the underlying secret sharing scheme, this verification method presents sub-exponential complexity, in a space versus time trade-off:
\begin{itemize}
\item{The time complexity of checking all $t$-subsets is $O(\binom {m} {t})$, which is super-polynomial in m.}
\item{Using the augmented histogram, also space complexity becomes $O(\binom {m} {t})$.}
\end{itemize}
However, in all practical applications of secret sharing, the maximum number of shareholders $n$, and therefore, $m$
and $t$, are values of order $10^1$, so the above considerations can be, in practice, disregarded.
\end{rem}

%%%%%%%%%%%%%%%%%%%%%%%%%%%%%%%%%%%%%%%%%%%%%%%%%%%%%%%%%%%%%%%%%%%%%%%%%%%%%%%%%%%%%%%%%%%%%%%%%%%%%%%%%%%%%%
\section{Space-efficient verifiability}
\label{schemes-section}
In this Section, we introduce our construction of a new verification method for threshold secret sharing. It is not designed for a particular scheme, nor does it require any assumption on the shares.
The designed verification algorithm is non-interactive (verification does not require receiving additional data from other shareholders, besides the shares), requires a one-time honest  dealer, and belongs to the family of commitment-based methods, since it relies on one-way functions. It will be shown that, under certain hypotheses, it is more space-efficient than the already illustrated homomorphic VSS extensions.

\subsection{Definitions}
Notations related to mathematical and string operators are listed below. The following convention will be used: any operator defined for a bitstring is valid for an unsigned integer type, and vice-versa.

\begin{itemize}
\item{$[s_1|s_2|\ldots|s_n]$ defines the concatenation of the bitstrings $s_1,s_2,\ldots,s_n$};
\item{$bs()$ denotes the bitsize of its argument. If the argument is an integer $n$, the bitsize is
$ bs(n) = 1+ \left\lfloor {{{\log }_2}n} \right\rfloor $.\\
If the argument is a set, the operator refers to the greatest element in the set:
$ bs(S)=bs(\max_{x \in S} x)$.\\Eg.: If $s = 11101_2$ and $S = \{5,7,111\}$, then $bs(s)=5$ and $bs(S)=bs(111)=7$};
\item $M(y)$ denotes the bitstring consisting of the most significant 
$\left\lceil {\frac{n}{2}} \right\rceil$ bits of the $n$-bit string $y$.
For example, if ${s = 11101_2}$, ${M(s)= 111_2}$. $L(y)$ denotes the bitstring consisting of the less significant 
$\left\lfloor {\frac{n}{2}} \right\rfloor$ bits of the bitstring $y$. Referring to the previous example, ${L(s)=01_2=1_2}$.\\
Clearly, for any string $s$, $s = [M(s)|L(s)]$ -- leading zeros in $L(s)$, if present, must be kept for a correct concatenation;
\item $NP(x)$ and $np(x)$ refer respectively to the lowest prime number strictly greater than $x$ and to the lowest prime greater than or equal to $x$.\\Eg.: $NP(22)=np(22)=23$, while $NP(11)=13$ and $np(11)=11$.
\end{itemize}

We recall here some definitions and useful results about permutations.

\begin{defn}[Permutation]
Given a set $I=\{1,\ldots,n\}$, a permutation over $I$ is a bijective mapping $\sigma:I \rightarrow I$. That is, every element of I maps to one (not necessarily different) element of $I$ itself, and no two different elements can map to the same one.
\end{defn}

\begin{lem}[Permutation over a probability distribution]~\\
\label{Lemma:permprob}
Let $\sigma: A \rightarrow B$ be a permutation, with $A = \{1,\ldots,n\}~,~B=\{\sigma(1),\ldots,\sigma(n)\}$; let $f_a:A \rightarrow [0,1]$ define a probability distribution\footnote{Probability Mass Function (PMF)} for the random variable $X_A$ over the set $A$, i.e.:
$$ P_A(X_A = i) = f_A(i) , i \in A $$
Then, the distribution obtained by applying the permutation $\sigma$ to the PMF $f_A$ is given by the set of probabilities that the random variable $X_B$ takes over the permuted items of the set B:
$$ \sigma(P_A(X_A=i)) = P_B(X_B = \sigma(i)) $$
\end{lem}

An immediate corollary of this is that the uniform distribution maps to itself under every possible permutation:
$$ \sigma(f_{\mathcal{U}}(i)) = f_{\mathcal{U}}(\sigma(i))~\forall \sigma:A\rightarrow B $$

\begin{lem}[Composition of permutations]
\label{Lemma:permcomp}
The space of permutation matrices of size n $(\Sigma^{n\times n})$ is a group under matrix product, hence permutations over input sets of equal size are closed under composition:

$$ \forall \sigma_i,\sigma_j \in \Sigma^{n\times n} , \sigma_i(\sigma_j) \in \Sigma^{n\times n} $$
\end{lem}

\begin{lem}
\label{Lemma:onewayperm}
Let $GF(q)$ be a finite field of prime size (not a polynomial field), $r$ one of its primitive roots, and $D = \{1,\ldots,q-1\}$. Then, the exponentiation function:
$$ e_r:D \rightarrow D~,~e_r(x) = r^x \mod q $$
is a permutation over D. % existence of fixed points: \citep{journals/corr/abs-1105-5346} -- simple example: 3^2 = 2 (mod 7) // 3^x mod 7 has three fixed points: 2,4,5
\end{lem}

The following result, related to the degree of an interpolating polynomial with regard to its interpolation points, will be used in the Powering polynomial (VSS-POW) scheme.

\begin{thm}
\label{thm:probdegreeinterp}
Let $(x_i,y_i)$, $i=1,\ldots,t$ be a set of $t$ random points with different abscissas $x_i$, and whose coordinates belong to a finite field $\mathbb{F}$ of prime cardinality $p$;
let $y(x) = \mathop \sum \limits_{i=0}^{t-1} {a_i x^i}$ be the interpolating polynomial of the given points, with coefficients over $\mathbb{F}$ as well.
Then, the probability that the degree of the polynomial  $y$ is strictly less than $t-1$ is negligible for big $p$:
$$ P[\deg(y) < t-1] = \frac{1}{p}$$
\end{thm}

\begin{IEEEproof}
Any set of points chosen following the given assumption, generates a full-rank Vandermonde matrix $X \in \mathbb{F}^{t\times t}$, which induces a bijection of the finite domain $\mathbb{F}^t$ onto itself:
$$\forall y \in \mathbb{F}^t~\exists!~a \in \mathbb{F}^t :~ Xa = y$$
for this reason, $X$ can be seen as a permutation of the elements of $\mathbb{F}^t$.
By Lemma \ref{Lemma:permprob}, the uniform discrete distribution is invariant with respect to permutations, so the probability of obtaining a polynomial of non-maximum degree -- with $a_{t-1}=0$ -- is equal to the one of choosing the $t$-th point with null ordinate\footnote{Notice that this \emph{does not} mean that a point with null ordinate generates a solution $a$ with a null coefficient, but that the cardinality of all points with the first property is equal to the one of polynomials with the second property; since domains coincide, probabilities are equal as well.
Also note that the point index is not relevant, $t$-th point has just been chosen in order to \emph{fix} a position, to distinguish from the case when any one of the points could have a null y-coordinate, which would lead to a wrong probability calculation.}:
$$ P[y(x) : a_{t-1} = 0] = P[y_t = 0] = \frac{1}{p} $$
\end{IEEEproof}

\subsection{Designing a space-efficient VSS extension}
The verification scheme that is going to be designed will be the result of incremental refinements of partially secure techniques. The main goal to achieve during the design will be the reduction of verification data. Labels of the form \emph{VSS-X} will be used to better identify and distinguish the variants obtained.
Moreover, since the final result is a commitment scheme, the security analysis will develop around the two security properties of hiding and binding.

\subsection{Security assumptions}

\begin{itemize}
\item There is a single, \emph{one-time}, \emph{honest} dealer, that distributes data to all of the $n$ shareholders involved in the scheme instance;
\item There is no trusted shareholder in the underlying network, and no storage of shared or public data. That is, once provided with their shares and verification data, shareholders do not need any other information for secret reconstruction and cheater identification;
\item Secure bidirectional channels can be established between pairs of entities - any external attacker can only be \emph{passive}, so man-in-the-middle attacks are not considered in this model; security against these kinds of attack is assumed to be addressed by the protocols that establish communication between the parties over a network (e.g., TLS);
\item Client machines are fully trusted. All of the entities (the dealer and the shareholders) run their respective protocol steps on their client machines where keys and certificates required for encryption/decryption and authentication are stored. If a CSP (Cloud Service Provider) has to be used for share storage, shareholders may encrypt their shares using a symmetric cipher before uploading them. Similarly, shareholders download shares from CSPs to their clients and decrypt them (if needed) before engaging in secret reconstruction and cheater identification; 
\item CSPs are semi-trusted and modeled as \emph{Honest-But-Curious} adversaries. Therefore, they act according to their prescribed actions in all of the protocols they are involved in (they do not, as \emph{malicious users} do, try to alter stored data and communications), but it is assumed that CSPs are interested in learning the contents of shares stored by shareholders, and can fully access everything stored on their cloud storage infrastructure. %A protocol is secure in the Honest-But-Curious model if and only if no player or coalition of $c$ < $n$ Honest-But-Curious players (who may cheat by sharing their private information) gains information about other players’ private input data, other than what can be deduced from the result of the protocol.
\end{itemize}

\subsection{Design features}
The main features our design attempts will insist on, are summarized below:
\begin{itemize}
\item \textbf{Commitments on shares:} Verification routines ensure that shares are legal \emph{independently} from the secret they are generated from, unlike homomorphic commitment schemes, that guarantee that a share corresponds to some secret;
\item \textbf{Non-interactivity:} Verification algorithms can be carried out in \emph{one interaction}, that is, no further communication with other parties is required after receiving the shares;
\item \textbf{Private verification:} each shareholder is able to verify the others' shares, but not its one: this is not necessary since this interaction model assumes a one-time honest dealer; moreover, verification is performed \emph{differently} by each shareholder, by taking as additional input a secret parameter.
\end{itemize}

\subsection{Powering polynomial (VSS-POW)}
Let $\mathcal{S}$ be a generic secret sharing scheme instance, with shares $s_i$, $i=1,\ldots,n$ belonging to some natural domain $D = \{0,1,\ldots,q\}$ and $bs(D)=bs(q)$ the domain's bitsize. The following is a non-interactive VSS extension based on polynomial interpolation. The dealer is in charge of doing the following steps:

\begin{itemize}
\item Choose a suitable finite field $\mathbb{F}$ for domain $D$: for example, $GF(NP(q))$ or $GF(2^{bs(q)})$;
\item Generate with Lagrange interpolation a polynomial $V(x)$ over $\mathbb{F}$ that maps the chosen shares to their powers with a random exponent $r \in \mathbb{F}$, and make it public to all of the shareholders, i.e.:

\begin{equation}
\label{vss-pow-verification}
V(s_i) = s_i^r , i=1,\ldots,n
\end{equation}

With high probability, $\deg(V)=n-1$ (see Theorem \ref{thm:probdegreeinterp}), so there will be $n$ coefficients to provide shareholders with.
\end{itemize}

A shareholder can verify the provided share by checking if it satisfies (\ref{vss-pow-verification}).

The bitsize of each coefficient is bounded by $bs(q)$, if the field chosen is $GF(2^{bs(q)})$, and by $bs(q)+1$ if $\mathbb{F}=GF(NP(q))$ (in the worst case, when $NP(q) \geq 2^{bs(q)}$).
Hence, like the hashing method, this approach suffers from a share expansion which is linear in the total number of shares generated. For example, by applying this VSS extension to a distributed-equations Shamir scheme with no public data, and considering as inputs to be verified $s_i = [x_i|y_i]$, the augmented share of each shareholder (reconstruction data + verification data) will be:
$$ x_i,y_i, v_{0},\ldots,v_{n-1} $$
where $v_i$ denotes the $i$-th coefficient of the verification polynomial. The total size is bounded by $bs(q)+n(bs(q)+1)$.

\subsubsection{Security analysis}
Theoretically, the proposed scheme could not be considered secure for hiding, in that shares can be discovered by finding the roots of the polynomial equation:
\begin{equation}
V(x)-x^r = 0
\end{equation}
Algorithms that are polynomial-time in the input polynomial's degree exist for this task, such as Berlekamp \cite{Berlekamp:1967:FPF}, Cantor-Zassenhaus, and Shoup \cite{Shoup90onthe}. For details about their asymptotic runtime complexity, see \cite{Shoup93factoringpolynomials}.
Notice however that, for VSS-POW, being the degree of $V(x)-x^r$ \emph{exponential} in the field bitsize, this scheme could be considered, on average, computationally hiding, if $r$ is chosen randomly.
In addition, it may happen that $V(x)=x^r$ admits other solutions than the actual shares: again, if they exist, they are found by factorization and extraction of linear factors, therefore binding is, at least, only computational.

\subsection{Verification by CRT solution is not efficient}
One may be tempted to try the same approach by using the solution of a remainder system, instead of the coefficients of a polynomial. For example, if the shares generated are $s_1,\ldots,s_n$, verification data could be a value $x$ such that:

$$ x \equiv \left\lfloor {\frac{s_i}{2}} \right\rfloor \pmod s_i $$

for every $i=1,\ldots,n$, and verification would be performed by checking that each share received satisfies the corresponding equation.
However, in order to have a unique solution, Chinese Remainder Theorem (CRT) requires the moduli of the system to be pairwise coprime: this imposes a restriction on the possible shares that can be verified.
It is not feasible as well to regenerate new shares until they are all coprimes among themselves, for two reasons:

\begin{itemize}
\item \textbf{Computational efficiency}: As a corollary of the Prime Number Theorem
\cite{books/daglib/0001130}, the probability that two integers sampled from the uniform discrete distribution $\mathcal{U}[2,N]$ are coprimes tends to $6/\pi^2$ as $N$ goes to infinity \cite{books/daglib/0083722}. This probability decreases super-polynomially as the number of values in which any pair should contain coprime numbers grows \cite{toth2002probability}. In principle, this could not really be a limitation, since a set of pairwise coprime numbers can be generated recursively starting with two numbers, and using trial and error methods together with repeated instances of the Greatest Common Divisor (GCD) algorithm;
\item \textbf{Space efficiency}: a worse issue prevents using the CRT-solution approach to obtain a VSS scheme: for what has just been stated, the density of sets of $n$ pairwise coprime values is very low for a given power set $\mathcal{P}$ over a domain, so the scheme instances would need an over-dimensioning in order to result secure to search attacks (when they are possible), in that an attacker would not need to check every possible set of integers, but only the groups of mutually coprime ones, which become very few with respect to the search domain as $n$ increases.
\end{itemize}

\subsection{String-split polynomial (VSS-SSP)}
The verifying-polynomial method introduced before can be modified in order to decrease the domain size of each coefficient, and so the maximum size of each verification element. Consider the complete set of shares $S=\{s_1,\ldots,s_n\}$ of a generic threshold scheme, with domain $D$ and share sizes $bs(s_i)\leq bs(D)$.
The distribution/verification algorithm under domain reduction should run as follows:\\

Dealer's steps

\begin{itemize}
\item Take the bitsize limit of any share, $bs(D)$,
and select the finite field for verification accordingly:
$\mathbb{F}=GF(NP(2^{\frac{bs(D)}{2}}))$
or $\mathbb{F}=GF(2^{\frac{bs(D)}{2}})$
;
\item For every share $s_i$, if $bs(s_i)<bs(D)$, obtain with zero-padding on the left the modified share $S_i$ such that $bs(S_i)=bs(D)$. Otherwise, let $S_i = s_i$;
\item Compute the two halves of the bitstring $S_i$ as ${S_{iM}=M(S_i)}$ and ${S_{iL}=L(S_i)}$;
\item Check that no two $S_{iM}$ are equal. If so, run the share generation algorithm again and go to the first step;
\item Interpolate the verification polynomial over $\mathbb{F}$ as $V(x)$ such that:
\begin{equation}
V(S_{iM})=S_{iL},~i=1,\ldots,n
\end{equation}
\item Broadcast the polynomial coefficients $v_0,\ldots,v_{n-1}$.
\end{itemize}

Upon receiving a share $s_i$, any shareholder can verify it by padding it to $S_i$ and checking if $V(S_{iM})=S_{iL}$.\\

\subsubsection{Security analysis: no binding}
This initial attempt is completely insecure against binding, since any one knowing $V$ can choose a random half string $a$, and provide the faked share $s' = [a|V(a)]$. However, it is a good starting point for reducing the size of verification data, and it can be made secure in combination with other approaches presented later.

\subsection{Enforcing binding: private verification}

Binding security can be enhanced, by making each of the $n$ shareholders verify the others' shares, and assigning a different, private security parameter $u_j$ to each verifier; in this environment, this would mean generating $n$ different polynomials, one for each shareholder, satisfying the equation:
\begin{equation}
V_j(x) = x^{u_j},~i=1,\ldots,n~,~i \neq j
\end{equation}
for the first method, or
\begin{equation}
V_j(M(x))=L(x)^{u_j},~i=1,\ldots,n~,~i \neq j
\end{equation}
for the second one. In other words, every shareholder would own a polynomial passing for the other shareholders' shares (or half-shares).

\subsubsection{Security analysis}
Possible attacks against VSS-POW with private verification are listed below:

\begin{itemize}
\item If $n-1$ organized cheaters conspire against the remaining shareholder, by applying polynomial GCD to ${V_1(x)-x^{u_1},}\ldots,V_{n-1}(x)-x^{u_{n-1}}$, they can obtain the missing share $s_n$, since every polynomial other than $V_n$ passes by $s_n$. Clearly, this attack does not result in a true gain, unless the threshold scheme to be protected is a $(n,n)$ one. We note here that this attack can be performed on the original VSS-POW scheme as well;
\item The same attack works for unhiding a missing share from a string-split polynomial. This time, after GCD, the factor retrieved will lead to $M(s_n)$, then the missing half-share can be retrieved by taking the $u_j$-th root of any available $V_j(M(s_n))$ -- efficient extraction of modular $n$-th roots can be performed using a generalization of the Tonelli-Shanks algorithm \cite{conf/focs/AdlemanMM77}.
\end{itemize}

However, like polynomial factorization algorithms, the polynomial variant of GCD requires polynomial-time \emph{in the input degree}, which is \emph{exponential} in the field's bitsize; therefore, this scheme is as secure against hiding as the one without private verification.

\subsection{Exponentiating polynomial (VSS-EXP)}
The variant that is going to be introduced now, will exploit some of the characteristics of the attempts made before, and a security assumption, in order to achieve computational security.\\

Let $s_1,\ldots,s_n$ be the input shares, and ${D=\{0,1,\ldots,q\}}$ and ${bs(D)=bs(q)}$ their domain and domain's bitsize, respectively. The dealer is in charge of doing the following steps:

\begin{itemize}
\item Choose a suitable finite field $\mathbb{F}$, such as $GF(NP(q))$ or $GF(2^{bs(q)})$;
\item For each shareholder $j$, select a primitive element $r_j$ of the multiplicative group $\mathbb{F}^*$, and generate with Lagrange interpolation a polynomial $V_j(x)$ over $\mathbb{F}$ that exponentiates all of the other shares through $r_j$\footnote{Notice that, while for prime order fields $GF(p)$ this equation is well-posed, for prime power fields $GF(p^k)$ we are performing a small abuse of notation: $s_i$ in the left-hand side of the formula is the element of the field (which is, actually, a polynomial), while the exponent on the right-hand side represents the natural number corresponding to the bitstring $s_i$, since, in finite field algebra, exponentiation by a polynomial is not defined.}:
\begin{equation}
\label{vss-exp-verification}
V_j(s_i) = {r_j}^{s_i},~i=1,\ldots,n~,~i \neq j
\end{equation}
\item Send $q$, $r_j$ and the coefficients of $V_j$ to shareholder $j$ via a secure private channel.
\end{itemize}

A shareholder can verify the provided share by checking if it satisfies (\ref{vss-exp-verification}).

\subsubsection{Security analysis}
The security of this scheme relies on the following assumption.

\begin{defn}[Exponentiating Polynomial Root Problem (EPRP)]~\\
Let $p(x)$ be a polynomial with ${\deg(p) \geq 0}$ with coefficients drawn from a finite field $GF(q)$, and $r$ a primitive element for that field. Then, the problem of finding roots of:
\begin{equation}
\label{eprp-equation}
p(x) = r^x
\end{equation}
is believed to be $NP$--intermediate, i.e., it is in the complexity class $NP$ but it is supposed not to be in $P$ nor $NP$--complete.
\end{defn}

It is worth noting that this problem is at least as hard as the DLP, in that it can be seen as a generalization of the latter -- DLP is the particular case of EPRP when $deg(p) = 0$ -- so a poly-time algorithm for solving EPRP would imply solution to any DLP instance. The problem is in $NP$, owing to the fact that, given a solution, verifying it consists in performing a number of modular additions, multiplications and exponentiations, which is linear in the number of coefficients; the runtime of this arithmetic is instead polynomial with respect to the field bitsize. To the best of our knowledge, as of this writing, efficient algorithms to solve this problem  do not exist. Finding roots to such equations can be done in two ways:
\begin{itemize}
\item Try all possible items $x$ in the field, and check whether they satisfy the equation or not. Clearly, even with randomized search, this requires \emph{exponential time} in the bitsize of the field modulus;
\item The exponential $r^x$ can be rewritten in polynomial form, by using Lagrange interpolation to interpolate the points $\{(0,r^0),(1,r^1),\ldots,({q-1},r^{q-1})\}$, determining a polynomial $f(x)$. This polynomial is \emph{identical} to $r^{x}$ precisely because we are working on a finite field. Then, the difference $p(x) - f(x)$, can be factored in order to find the roots of the given equation (using Berlekamp, Cantor--Zassenhaus or Shoup algorithms) and the roots read off the factors. However, this approach is even worse than exhaustive search: since, on average, a polynomial passing by $n$ given points will have $n$ non-null coefficients, even only the input to Lagrange interpolation will require exponential space in the field bitsize. Also, notice that, in this case, there is no space-time tradeoff: Lagrange interpolation is an algorithm that uses \emph{entirely} its input: this means that, for instances with inputs requiring exponential space, runtime would be exponential as well.
\end{itemize}

The security of this scheme can be summarized as follows:
\begin{itemize}
\item \textbf{Reverse hiding:} A dishonest shareholder willing to obtain all of the others' shares from his verification polynomial, should try, on average, about $2^{bs(q)}$ values\footnote{Given a domain of $N$ strings, $k$ of which representing valid shares, the expected number of trials for the $k$--th success when sampling without replacement (i.e. finding all of the shares via a randomized exhaustive brute-force attack), is $k(N+1)/(k+1)$, a value rapidly approaching $N$.}.
Also, notice that the equation may present additional roots other than the valid shares: if this happens, an unbounded adversary could exploit the set coherence method (\ref{setcoherence}), and obtain the secret as the majority value, without caring about which solutions are legal or not;
\item \textbf{Reverse binding:} since no one except shareholder $j$ knows the primitive element $r_j$ used in the construction of $V_j$, in order to be able to deceive a verification equation, $r_j$ must be guessed, and the equation must present additional solutions. For commitments in which the equation has only roots in the valid values, binding is perfect.
\end{itemize}

\subsection{String-split exponentiating polynomial (VSS-EXP-SSP)}
The string-split approach can be applied to VSS-EXP to reduce the total amount of data targeted to each shareholder. It will be proved that, by keeping the assumption made, security of this scheme is equivalent to the original one's. The dealer is in charge of doing the following steps:

\begin{itemize}
\item Given the set of shares $S$, choose a suitable field $\mathbb{F}$ for half-shares $M(s_i),L(s_i)$ as $GF(2^w)$ or $GF(NP(2^w))$, with $w = \left \lceil \dfrac{bs(S)}{2} \right \rceil$ an upper bound on the shares' half sizes;
\item For each shareholder $j$, select a primitive element $r_j$ of $\mathbb{F}^*$, and generate with Lagrange interpolation a polynomial $V_j(x)$ over $\mathbb{F}$ such that:
\begin{equation}
\label{vss-exp-ssp-verification}
V_j(M(s_i)) = {r_j}^{L(s_i)},~i=1,\ldots,n~,~i \neq j
\end{equation}
\item Send $w$, $r_j$ and the coefficients of $V_j$ to shareholder $j$ via a secure private channel.
\end{itemize}

A shareholder can verify the provided share by checking if it satisfies (\ref{vss-exp-ssp-verification}).\\

\subsubsection{Security equivalence}
This scheme has the same security properties of VSS-EXP, under the same assumptions.

\begin{thm}
Let $\mathcal{V}_1$ and $\mathcal{V}_2$ be two
instances of \begin{small}VSS-EXP\end{small} and
\begin{small}VSS-EXP-SSP\end{small}
respectively, with the shares domain bitsize
for $\mathcal{V}_1$ being half the one for
$\mathcal{V}_2$.
The two instances are stochastically equivalent
for hiding and binding.
\end{thm}

\begin{IEEEproof}
Shares can be considered random bitstrings sampled from the domain $\{0,1\}^b$. A random bitstring of size $b$ -- suppose $b$ even, without loss of generality -- can be seen as concatenation of two random bitstrings of size $b'=\frac{b}{2}$.
Suppose $n$ $b$-bitstrings are chosen uniformly; the probability of not extracting the same string twice is:
$$ \left(\frac{2^b-1}{2^b}\right) \left(\frac{2^b-2}{2^b}\right) \ldots \left(\frac{2^b-n+1}{2^b}\right) = \dfrac{(2^b-1)!}{(2^b-n)!2^{b(n-1)}}$$
Analogously, the probability of choosing $n$ $b$-bitstrings $s_i$ such that no two $M(s_i)$ are equal, and, independently, no two $L(s_i)$ are equal, is given by:
$$ \left(\dfrac{(2^{\frac{b}{2}}-1)!}{(2^{\frac{b}{2}}-n)!2^{{\frac{b}{2}}(n-1)}}\right)^2 $$
Clearly, fixing $n$, both probabilities approach $1$ as $b$ grows large.

Since both strings and half-strings are supposed to be extracted from a random uniform process, for proving stochastic equivalence it is sufficient to prove that solving the VSS-EXP equation:
\begin{equation}
V(x) = r^x
\end{equation}
is statistically as hard as solving the equation:
\begin{equation}
\label{stat-equivalent}
V(x_1) = r^{x_2}
\end{equation}
where $x,x_1,x_2$ belong to the same domain $D$ (or at least to domains with the same bitsize).

Indeed, this is true for the following reasons:

\begin{itemize}
\item Suppose that all of the $x_1$ and, independently, all of the $x_2$ values chosen for interpolation are different: this is practically always true, given the probabilities defined before; then, a permutation ${\sigma_1:D \rightarrow D}$ exists, mapping each $x_1$ to one and only one $x_2$;
\item By Lemma \ref{Lemma:onewayperm}, exponentiation $r^x$ defines another permutation ${\sigma_2:D\rightarrow D}$, if we exclude from $D$ the value $0$; again, the probability of extracting at least one $0$ value for any $x_1$ or $x_2$ is negligible, if $b$ is sufficiently large;
\item (\ref{stat-equivalent}) can be rewritten as:
\begin{equation}
V(x_1) = r^{\sigma_1(x_1)} = \sigma_2(\sigma_1(x_1))
\end{equation}
Since, from Lemma \ref{Lemma:permprob}, the uniform distribution holds invariance with respect to permutations (and also compositions of permutations, by closure of the permutation group -- Lemma \ref{Lemma:permcomp}), the equation of VSS-EXP presents an equivalent distribution of solutions of VSS-EXP-SSP, provided input domains are equal or similar in size. Then, if the EPRP assumption is valid, the two schemes are cryptographically equivalent.
\end{itemize}
\end{IEEEproof}

\section{Runtime efficiency refinements}
\label{runtime-efficiency-section}
The VSS-EXP family requires computation of random primitive elements, in order for exponentiation to span over the whole multiplicative group of interest. The efficiency refinements presented here will refer to prime order fields $GF(p)$. Some special cases of comfortable prime power fields of binary form, $GF(2^n)$, will be presented later.

Given a prime $p$ and the modular multiplicative group ${\mathbb{Z}_p^* = \{1,\ldots,p-1\}}$, a primitive root for that group is a generator whose order is $p-1$.
Since no efficient algorithms exist for finding primitive roots modulo a prime, random trial-and-error methods are used:

\begin{itemize}
\item Choose a random number $r$ from the uniform distribution ${\{2,\ldots,p-1\}}$; note that $1$ is only a generator of the trivial group $\{1\}$, since $1^x$ is always $1$ for any $x$, so it can never be a primitive root for non-trivial groups;
\item Compute the multiplicative order of $r$: if it is equal to $p-1$, stop; otherwise, go to the previous step.
\end{itemize}

However, even computing multiplicative orders is, in general, a hard problem: 

\begin{itemize}
\item Any number $a$ in the multiplicative group $\mathbb{Z}_p^*$ must have as order a divisor of $p-1$; so, the standard trial-and-error technique here consists in evaluating ${a^d \mod p}$ for all of the divisors of $p-1$, and taking as result the minimum argument $d$ for which $a^d \equiv 1 \pmod p$;
\item If $p-1$ is hard to factor, for example, if $p-1 = qs$, with $q,s$ being large primes, then it is also hard to compute orders.
\end{itemize}

Hence, in order to efficiently compute primitive roots, one should choose the field modulus $p$ for the verification polynomials, such that $p-1$ is easy to factor. One such way is choosing $p-1$ as a smooth number (i.e. a number that factors into small primes); however, notice that efficient computation of discrete logarithms can be carried out in a multiplicative group of smooth size, thanks to the Silver-Pohlig-Hellman algorithm \cite{Mollin:2008:FNT:1628707}.

\begin{defn}[Safe primes, Sophie Germain primes]
Let $p$ be a prime number; $p$ is \emph{safe}, if $\frac{p-1}{2}$ is also prime.
Conversely, a prime $q$ is a Sophie Germain prime, if $2q + 1$ is also prime.
\end{defn}

The number $\pi_{sg}(x)$ of Sophie Germain primes less than a given $x$ (or equivalently, of safe primes less than $2x$) has been conjectured \cite{books/daglib/0018101} to be

\begin{equation}
\pi_{sg}(x) = \frac{Cx}{\left(\ln x\right)^2},~C \simeq 1.32032
\end{equation}

\subsection{Advantages of choosing a safe prime as modulus}
The are some good reasons for working in a field having a safe prime as modulus:
\begin{itemize}
\item \textbf{Order computation}: if $p$ is safe, $p-1 = 2q$, then any number $a$ of the multiplicative group $\mathbb{Z}_p^*$ can have as order $2$, $q$, or $p-1$. Hence, at most $2$ exponentiations have to be performed to compute an order -- for $p$ prime, $a^{p-1}$ is always $1$, by Fermat's theorem \cite{books/daglib/0001130};
\item \textbf{Number of primitive roots}: the number of primitive roots in $\mathbb{Z}_p^*$ with $p$ safe, is:

\begin{IEEEeqnarray}{rCl}
\phi(\phi(p)) &=& \phi(p-1)=\phi(2q)  \IEEEnonumber\\
 &&=\ \phi(2)\phi(q)=q-1  \IEEEnonumber\\
 &&=\ \frac{p-1}{2} -1%
\end{IEEEeqnarray}

So, by random sampling, one expects to find, on average, a primitive root after $2$ attempts.
Even better, since any primitive root $g$ modulo $n$ generates all of the other ones as:
\begin{equation}
g^a \mod n,~\gcd(a, \phi(n))=1
\end{equation}
it is enough to choose one primitive root -- for example, the lowest one -- and then compute the others with random values $a$ coprime to $p-1$ and $2$, i.e.: ${a \in \{3, 5, 7, ,\ldots,p-2\}/\{q\}}$.
\end{itemize}

Summing up, a VSS-EXP scheme exploiting safe primes should work as follows:

\begin{itemize}
\item Choose as finite field $\mathbb{F}$, $GF(p)$, with $p=NSP(q)$ the next safe prime greater than $q$; the safe prime can also be chosen as ${p=NSP(2^{bs(q)})}$;
\item For each shareholder $j$, select a primitive root $r_j$ of $\mathbb{F}$, and generate with Lagrange interpolation a polynomial $V_j(x)$ over $\mathbb{F}$ that exponentiates all of the other shares through $r_j$:
\begin{equation}
\label{vss-exp-verify}
V_j(s_i) = {r_j}^{s_i}~,~i=1,\ldots,n~,~i \neq j
\end{equation}
\item Send $p$, $r_j$ and the coefficients of $V_j$ to shareholder $j$ via a secure private channel;
\item A provided share is verified by checking if it satisfies (\ref{vss-exp-verify}).
\end{itemize}

\subsection{Prime power fields from Mersenne primes}
There are some special cases of prime power fields ${\mathbb{F}=GF(2^n)}$, for which order computation is not needed.
\begin{lem}
Let $p$ be the exponent of some Mersenne prime $2^p-1$. Then, the multiplicative group $\mathbb{F}^*$ of the finite field $\mathbb{F}=GF(2^p)$ contains only primitive elements, except $1$.
\end{lem}
\begin{IEEEproof}
Since the size of the group is a prime number, no element can have an exponentiation period lower than $2^p -1$, so every element greater than $1$ in the field is primitive.
\end{IEEEproof}

%%%%%%%%%%%%%%%%%%%%%%%%%%%%%%%%%

\begin{rem}
If DLP and EPRP are polynomially equivalent, or computationally related, working in groups of smooth cardinality would result in a loss of security, since an efficient discrete logarithm computation would lead to efficient root extraction for the exponentiating polynomial. Instead, using safe primes of high Hamming weight \footnote{The Hamming weight of a bitstring is the number of its bits set to 1.} \footnote{For low Hamming weight safe prime moduli, a specialized algorithm, SNFS -- Special Number Field Sieve --, can compute discrete logarithms more efficiently than in the general case.} would remain a good choice, since the derived groups are not suitable -- at least as of this writing -- for efficient logarithm computation.
With random search, safe primes up to $2048$ bits can be found in a few minutes on modern CPUs. Moreover, lists of bigger safe primes are publicly available online, for example the one in \cite{rfc3526}.
\end{rem}

\begin{rem}
Computation of primitive elements in prime power
fields $GF(p^k)$ requires finding a primitive polynomial
over $GF(p)$.
A list of primitive polynomials for
binary fields $GF(2^k)$ up to degree $k=5000$
(and, in particular, for Mersenne exponents in that range)
is given in \cite{Zivkovic:1994:TPB:179653.179706}.
\end{rem}

\section{Information rates}
\label{comparison-section}
In this Section we discuss the amount of verification data sent to each shareholder by the dealer for both the VSS-EXP and VSS-EXP-SSP schemes, and compare them against Feldman's scheme.

\subsection{VSS-EXP}

\begin{itemize}
\item Public parameters: $q$ ($p_1=NSP(q)$ is uniquely determined) or $bs(q)$, if $p_1=NSP(2^{bs(q)})$ --- $bs(bs(q))$ bits;
\item Private security parameter: $r_j$ -- at most $bs(p_1)$ bits;
\item Polynomial coefficients: at most $(n-1) \cdot bs(p_1)$ bits.
\end{itemize}

\subsection{VSS-EXP-SSP}

\begin{itemize}
\item Public parameters: $w$ ($p_2=NSP(2^w)$ is uniquely determined) -- $bs(w)$ bits;
\item Private security parameter: $r_j$ -- at most $bs(p_2)$ bits;
\item Polynomial coefficients: at most $(n-1) \cdot bs(p_2)$ bits.
\end{itemize}

The total amount of bits is then limited by:
\begin{equation}
bs(bs(q))+n\cdot bs(p_1)
\end{equation}
for VSS-EXP and
\begin{equation}
bs(w) + n\cdot bs(p_2)
\end{equation}
for VSS-EXP-SSP.

\subsection{Comparison with other commitment-based schemes}
In the following, we compare our schemes against Feldman's scheme  (\ref{feldmanvss}). We do not take into account Pedersen's (\ref{pedersenvss}), owing to the fact that, as already discussed, its verification data is bigger than Feldman's. Moreover, we do not compare our schemes against Benaloh's (\ref{benalohvss}), since that scheme requires a huge number of polynomials, corresponding to a lot of verification data. We also disregard set coherence (\ref{setcoherence}) since it requires for a $(t, n)$ threshold scheme a coalition consisting of $m > t$ shareholders, and cheater detection/identification is performed by comparing the secrets reconstructed by all of the possible subsets of $t$ out of $m$ shareholders, thus requiring a greater number of reconstruction operations.  

The scheme proposed by Feldman outputs the following verification data, taking as input a polynomial $y(x)$ with $t$ coefficients from $GF(q)$, of maximum size $bs(q)$ each:

\begin{itemize}
\item Two public parameters $p,q$, whose size depends on the computational effort needed to solve a DLP instance: as of this writing, $p$ should be at least $2048$ bits long;
\item $t$ commitments in $\mathbb{Z}_p^*$, each one of maximum bitsize $bs(p)$.
\end{itemize}

The total amount of data that each shareholder must know in order to verify a share is:
\begin{equation}
bs(p) + bs(p) t  = (t+1)bs(p) = K(t+1)bs(q)
\end{equation}
where $K=\frac{bs(p)}{bs(q)}$ denotes the bit expansion factor from $\mathbb{Z}_q$ to $\mathbb{Z}_p$.
Note that $q$ is not considered in the summation, since this datum is a parameter of the Shamir's scheme instance to be verified, and not properly of the VSS scheme. %K=8

The information rate for a commitment scheme is computed as the ratio between the total size of commitments, and the quantity of data for which they are computed. For Feldman's scheme, it is:
\begin{equation}
R = \frac{bs(p)(t+1)}{{bs(q)}t} = K \frac{t+1}{t}
\end{equation}
This rate varies depending on the secret's size: for example, for a $160$-bit secret, $K = \frac{2048}{160} = 12.8 $, so the rate would be very high; a better rate can be obtained with longer secrets, and higher threshold values.

Regarding the VSS-EXP family, the total amount of data to be committed depends on the number of shareholders $n$ in a secret sharing session; assuming shares in $GF(q)$, this quantity is limited by $bs(q) n$ bits. The rates can then be computed as:
\begin{equation}
R_{_{VSS\_EXP}} = \frac{bs(p_1)}{bs(q)}
\end{equation}
and
\begin{equation}
R_{_{VSS\_EXP\_SSP}} = \frac{bs(p_2)}{bs(q)} \simeq \frac{bs(p_1)}{2 bs(q)}
\end{equation}

Clearly, since, as of this writing, no better methods than exhaustive search are known to solve EPRP, no additional lower bounds are imposed on the size of committed data for VSS-EXP, so that the computed rates should result, in general, much lower than the DLP-based commitments counterparts, at least for moderate secret sizes ($128\sim 256$ bits).

\section{Conclusions}
\label{conclusions-section}
We have presented new verification extensions enhancing arbitrary secret sharing schemes by adding cheater detection capabilities. Our main effort was devoted to reducing the amount of verification data for a secret sharing scheme without worsening the security properties; a new computational problem, EPRP, supposed to be harder than the DLP, has been introduced, but the derived verification schemes, missing the homomorphic property, are not extensible to additional shareholders, and the dealer must be a trusted entity, since any malicious behaviour of this party cannot be detected.
Further research should be carried out on the possibility of modifying the proposed problem in order to augment it with the homomorphic property, so that a resulting VSS scheme would present shareholder extensibility, and to investigate if this kind of problem can be also exploited in interactive proofs for authenticating the dealer's integrity and in public-key based cryptosystems.  Another possible direction for future work could regard investigating additional runtime efficiency refinements. Finally,  proving the NP-hardness of EPRP by deriving a suitable poly-time reduction would result in a substantial breakthrough in computer science.

\bibliographystyle{ieeetr}
\bibliography{bibliography}

% biography section
% 
% If you have an EPS/PDF photo (graphicx package needed) extra braces are
% needed around the contents of the optional argument to biography to prevent
% the LaTeX parser from getting confused when it sees the complicated
% \includegraphics command within an optional argument. (You could create
% your own custom macro containing the \includegraphics command to make things
% simpler here.)
%\begin{IEEEbiography}[{\includegraphics[width=1in,height=1.25in,clip,keepaspectratio]{mshell}}]{Michael Shell}
% or if you just want to reserve a space for a photo:

\begin{IEEEbiography}[{\includegraphics[width=1in,height=1.25in,clip,keepaspectratio]{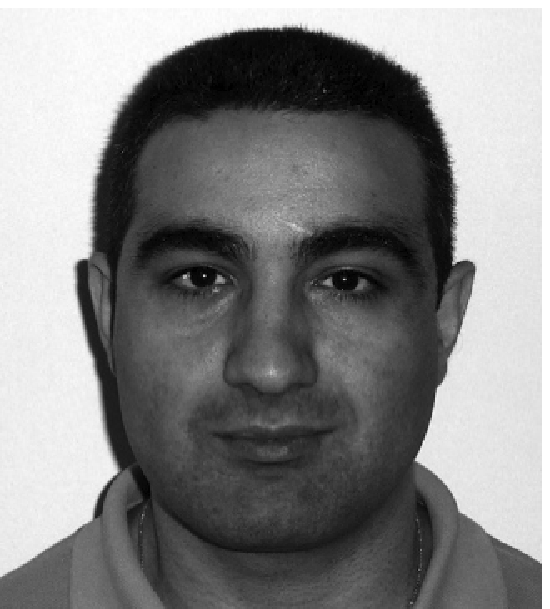}}]{Massimo Cafaro}
is an Assistant Professor at the Department of Innovation Engineering of the University of Salento. His research covers High Performance, Distributed and Cloud/Grid Computing, security and cryptography. He received a Laurea degree in Computer Science from the University of Salerno and a Ph.D. in Computer Science from the University of Bari. He is a Senior Member of IEEE and of IEEE Computer Society, and Senior Member ACM. He authored or co-authored more than 90 refereed papers on parallel, distributed and grid/cloud computing. He co-authored and holds a patent on distributed database technologies. His research interests are focused on both theoretical and practical aspects of parallel and distributed computing, security and cryptography, with particular attention to the design and analysis of algorithms.
\end{IEEEbiography}

\begin{IEEEbiography}[{\includegraphics[width=1in,height=1.25in,clip,keepaspectratio]{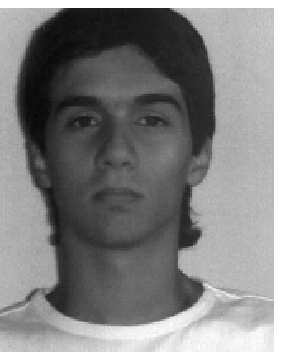}}]{Piergiuseppe Pell\`e}
received the M.Sc. degree in Computer Engineering from the University of Salento. His interests are in the field of security and cryptography.
\end{IEEEbiography}

% You can push biographies down or up by placing
% a \vfill before or after them. The appropriate
% use of \vfill depends on what kind of text is
% on the last page and whether or not the columns
% are being equalized.

%\vfill

% Can be used to pull up biographies so that the bottom of the last one
% is flush with the other column.
%\enlargethispage{-5in}

% that's all folks
\end{document}